\documentclass[twocolumn,amsmath,amssymb,pra,showpacs,superscriptaddress,aps]{revtex4}

\usepackage[english]{babel}
\usepackage[applemac]{inputenc}
\usepackage[T1]{fontenc}
\usepackage{verbatim}
\usepackage[pdftex]{graphicx}
\usepackage[normalem]{ulem}
\usepackage{color,braket}

\definecolor{dgreen}{rgb}{0.0, 0.7, 0.0}
\definecolor{mycolor}{rgb}{0.0, 0.3, 0.0}

\begin{document}

\bibliographystyle{aip}

\title{Generation of robust entangled states in a non-hermitian 
periodically driven two-band Bose-Hubbard system}

\author{Carlos A. Parra-Murillo}
\affiliation{Departamento de F\'isica, Universidad Del Valle, A. A. 25360,Cali, 
Colombia}
\author{Manuel H. Mu\~noz-Arias}
\affiliation{Departamento de F\'isica, Universidad Del Valle, A. A. 25360,Cali, 
Colombia}
\author{Javier Madro\~nero}
\affiliation{Departamento de F\'isica, Universidad Del Valle, A. A. 25360,Cali, 
Colombia}
\affiliation{Centre for Bioinformatics and Photonics-CIBioFi, Calle 13 No. 100-00, Edificio 320, No. 1069, Cali, 
Colombia}
\author{Sandro Wimberger}
\affiliation{DiFeST, Universit\`{a} degli Studi di Parma, Parco Area delle Scienze 7/a, 43124 Parma, Italy}
\affiliation{INFN, Sezione di Milano Bicocca, Gruppo Collegato di Parma, Parma, Italy}
\affiliation{ITP, Heidelberg University, Philosophenweg 12, 69120 Heidelberg, Germany}

\begin{abstract}
A many-body Wannier-Stark system coupled to an effective reservoir is studied within a non-Hermitian approach in the presence of a periodic driving. We show how the interplay of dissipation and driving dynamically induces a subspace of states which are very robust against dissipation. We numerically probe the structure of these asymptotic states and their robustness to imperfections
in the initial-state preparation and to the size of the system. Moreover, the asymptotic states are found to be strongly entangled making them interesting for further applications.
\end{abstract}

\date{\today} 

\maketitle

\section{Introduction}
\label{sec:1}

Experiments with ultracold atoms \cite{BDZ2008, greiner1, greiner2} give direct access to many-body quantum systems that often cannot be described theoretically due to the enormous computational complexity \cite{Qsim}. Quite challenging in this respect are many-body models either in more dimensions \cite{2D, kolovskypra, FHW2016} or with more modes per lattice site \cite{Bands, FW2016}, extending e.g. single-band Hubbard models \cite{JZ1998}.  

The coupling of energy bands in tilted Wannier-Stark systems was demonstrated experimentally \cite{Pisa, AW2011}. The effect of many-body interactions on the interband coupling was studied theoretically to great detail, in the weakly interacting limit \cite{ploetz} as well as in the context of strongly correlated Bose-Hubbard models \cite{tomadin, parraphd, Comp2015}. The latter many-body models, however, restricted to two bands only and neglected the opening of the system arising from the coupling to higher lying energy bands. 

In this paper, we discuss an open two-band Bose-Hubbard model with an additional periodic driving force to control the coupling between the two bands. Similar Floquet engineering of ultracold quantum gases are discussed, e.g., in \cite{Pisa-D,Innsbruck2016}. We include the loss channel to higher bands using a non-hermitian Hamiltonian approach \cite{nonhermitian}. While this method is valid only for small loss, it still allows us the access to the Floquet spectrum of the periodically driven system. In particular, we show that the simultaneous presence of dissipation and quasi-resonant driving between the two band induces the formation of states with interesting properties. First these states are extremely robust with respect to dissipation, and second they correspond to highly entangled superpositions of Fock states. 

The paper is organized as follows. Section \ref{sec:2} introduces the system we investigate as well as the origins of its temporal dependence. Subsection \ref{sec:2-2}, in particular, discusses the manifold of states coupled in our system. The dynamics induced by the dissipation and its application in the generation of highly entangles states are reported in section \ref{sec:3}. Section \ref{sec:concl} then concludes the paper.
 
\section{Non-Hermitian approach and Bose-Hubbard Hamiltonian}
\label{sec:2}

This section introduces our many-body Bose-Hubbard model with two coupled energy bands, external driving and the dissipation process. First we discuss the general properties of the system, before addresses the specific structure of states participating in the dynamical evolution.

\subsection{Many-Body Wannier-Stark System}
\label{sec:2-1}

The Wannier-Stark (WS) problem in its simplest form consists of a quantum particle trapped 
in a one-dimensional periodic potential, $V(x+d_L)=V(x)$, subjected to 
an additional static force of magnitude $F$. The Hamiltonian of this 
system, $\hat H_0 = \hat p^2/2m+V(x)+Fx$, is analytically diagonalized within the tight-binding 
approach. The eigenenergies, $E_l=l d_L F$, form the so-called WS ladder 
responsible for the occurrence, for instance, of Bloch oscillations \cite{kolovskyphysrep}, see e.g. 
\cite{kol68PRE2003, Innsbruck2008} for many-body studies of them and \cite{AW2011, Innsbruck2013, Innsbruck2014} for related
 resonant tunneling phenomena.
Here, $d_L$ is the lattice constant and $l$ is an integer indexing a single well or lattice site of the periodic 
potential $V(x)$. The WS eigenstates are embedded resonances  
within the continuous unbounded quantum spectrum of $\hat H_0$.  
Therefore the WS eigenstates are metastable states with 
lifetimes $\tau_l\sim 1/\Gamma_{l}$. These lifetimes are associated with 
the imaginary part of the corrected single-particle eigenenergies $\varepsilon_l= E_l-i\Gamma_{l}/2$,
see, e.g., Ref.~\cite{kolovskyphysrep} for details on the single-body case. 

\begin{figure}[t]
\includegraphics[width=0.95\columnwidth]{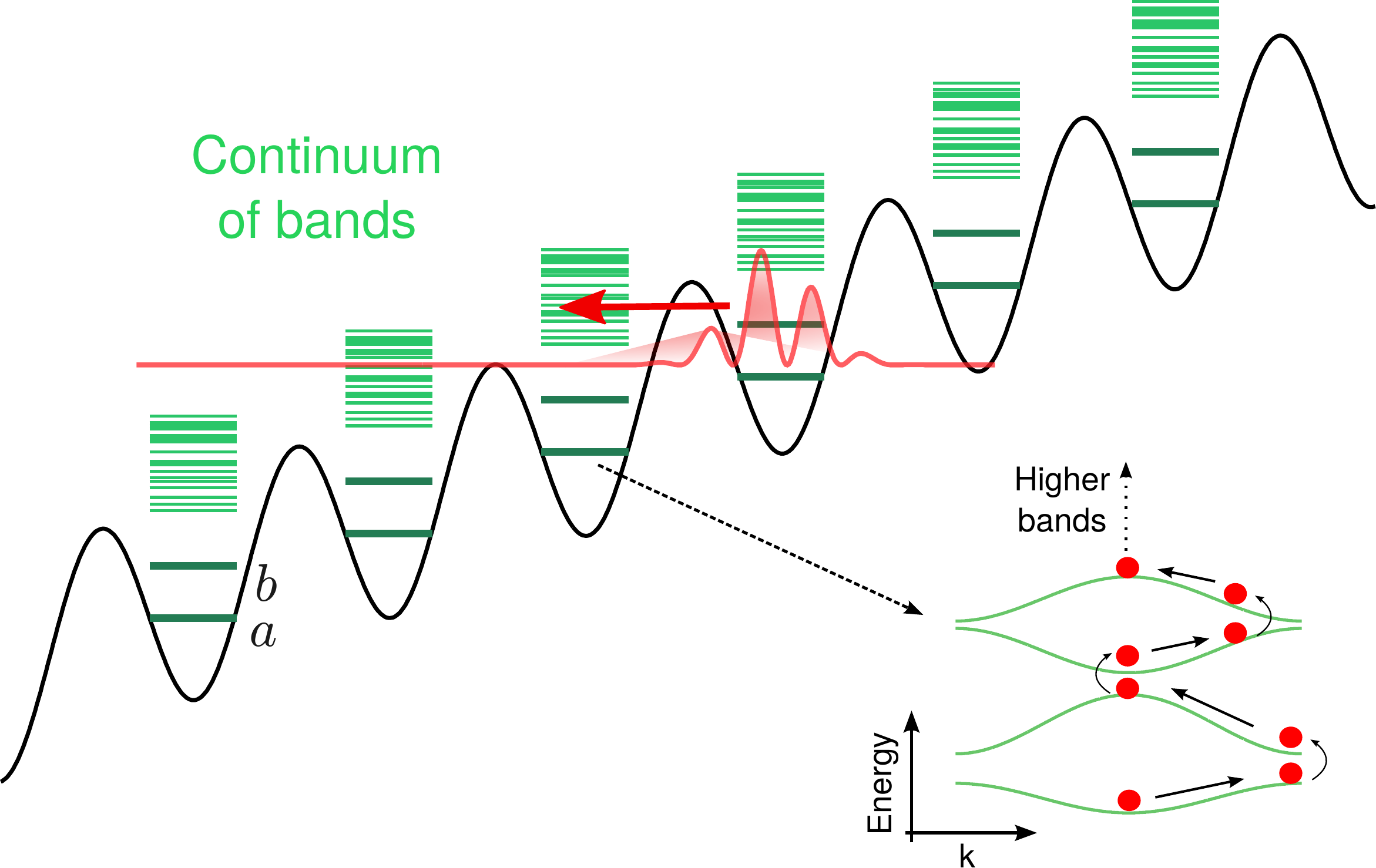}
\caption{\label{fig:intro} ({\em Color online}) Scheme of the 
single-particle Wannier-Stark ladder. The bold lines correspond
to the two-lowest (quasi)bound states (marked by $a$ and $b$) considered in this paper.
The structure can be understood as the remainders of the Bloch bands 
after turning on the external field. The wave packet (red) 
represents a Wannier-Stark state that leaks to the continuum of higher bands.
This process is marked by the arrow going to the left in the upper panel.} 
\end{figure}

Our system of interest is a many-body version of the WS Hamiltonian ultimately
constrained to the two-lowest WS ladders, see the sketch in Fig.~\ref{fig:intro}. 
The respective Hamiltonian in second quantization can be build following 
the standard procedure
\begin{eqnarray}\label{eq:01}
\hat H &=& \int dx\, [\hat\phi^{\dagger}(x) \hat H_0\hat\phi(x)+
g\hat\phi^{\dagger}(x)\hat\phi^{\dagger}(x)\hat\phi(x)\hat\phi(x)],\;\;\;\;\;\;
\end{eqnarray}
with the field operators defined in terms of site-localized Wannier functions of the 
flat lattice, i.e. for $F=0$: $\hat\phi(x) = \sum_{\alpha,l} \chi_l^{\alpha}(x)\hat\alpha_l$. 
$\alpha$ is the Bloch band index and $l$ is the site index. The 
Hamiltonian might be seen as an infinite 
collection of tilted single-band Bose-Hubbard Hamiltonians, plus additional coupling terms between them. 
The coupling is mainly induced by the external force $F$, but also by the interparticle interaction characterized by the
constant $g$ \cite{parraphd}. Clearly, the force breaks the translational 
invariance 
of the flat lattice, therefore the Bloch band picture interpretation strictly speaking fails. 
Instead, we have WS ladders (see Fig.~\ref{fig:intro}) and an entire zoo of single-
and many-body processes to be identified and investigated in detail, see e.g. \cite{parraphd,ploetz,tomadin, Comp2015}. 

%%% In order to treat the many-body WS problem, the extended Hilbert space
Generally, the total state space is spanned by the following Fock states
$\{|n^{\alpha_1}_1 \ n^{\alpha_1}_2\cdots, n^{\alpha_2}_1 \ n^{\alpha_2}_2 
\cdots, \ldots, n^{\alpha_k}_1 \ n^{\alpha_k}_2\cdots\rangle\}$, where $k\in 
\mathbb N$ is the number of bands taken into account. The integers 
$n^{\alpha_i}_l$ denote the number
of atoms in the potential well $l$  
and band $i$.

In what follows we restrict to only the two lowest WS ladders $\alpha_{j=1,2}$. 
The remaining ladders $\alpha_{j>2}$ act as an effective 
environment forming a continuum of bands (Fig.~\ref{fig:intro}). The system
dynamics is studied by means of the reduced density operator 
$\hat\rho_S={\rm Tr_{\alpha_{j>2}}}[\hat\rho]$. The evolution of $\hat\rho_S$ is
approximately described by a markovian quantum master equation of the form 
\begin{eqnarray}\label{eq:02}
  \partial_t\hat\rho_S=-i(\hat H_{\rm eff}\hat\rho_S-\hat\rho_S \hat H_{\rm eff}^{\dagger})
  +\sum_{\alpha,l} \hat C_l^{\alpha}\,\hat\rho_S\, \hat C_l^{\alpha\dagger}.
\end{eqnarray}
$\hat  H_{\rm eff} = \hat H - i\sum_{\alpha,l}  \hat C_l^{\alpha \dagger}\hat C_l^{\alpha}$ is an
effective non-Hermitian Hamiltonian defined by the quantum jump operators $\hat C_l^{\alpha}$ 
and the isolated-system Hamiltonian $\hat H$. The jump operators $\hat C_l^{\alpha}=\sqrt{\gamma_{l}^\alpha} \hat \alpha_l$ 
describe the particle loss to the higher bands. 

We use an approximate non-hermitian Hamiltonian approach to address our problem. 
This method has the advantage 
that we have access to the spectrum of the effective Hamiltonian without having  
to solve the
master equation for the reduced density matrix. The method consists in omitting the last term on the right-hand-side of 
Eq.~\eqref{eq:03}. The imaginary parts of the effective Hamiltonian give diagonal contributions in the Fock basis, which are proportional to the occupation numbers in the upper-WS ladder. In the end, we obtain a new effective Hamiltonian which is non-hermitian, but can 
simply be diagonalized in the Fock space of fixed particle number. This 
dramatically simplifies the numerical effort. The non-hermitian part
of the effective Hamiltonian nonetheless reduces the norm of the evolving states and  induces a dynamical dephasing 
via the decay rates $\gamma_{l}^\alpha$. The approach is valid for $\gamma_{l}^{\alpha}$ are 
much smaller than the relevant energy scales of the system. Hence, here we use rates $\gamma_{l}^{\alpha}\ll |J_{\alpha}|$. 
Following Ref.~\cite{parraphd}, we can argue, that in a doubly periodic optical lattice, we obtain
situations in which the two lowest (mini)bands are sufficiently isolated from the upper bands, such that the assumptions
just discussed are indeed fulfilled. Furthermore, we assume that only the 
upper band is effectively coupled to the higher bands.

%. This latter consists in analyzing the dynamics 
%only considering the Lioville-like equation defined for non-Hermitian Hamiltonian
%$\hat H_{\rm eff}$ and omitting last right-hand term (r.h.s) in Eq.~\ref{eq:03}.
%This approach is well suited if the decay rates in the Linbland term 
%$\gamma_{l}^{\alpha}$ are much smaller than the relevant energy scales of the 
%system. Here we set $\gamma_{l}^{\alpha}\ll |J_{\alpha}|$. Following Ref.~\cite{parraphd}
%we consider a double-well periodic optical potential which helps to isolate the two lowest 
%Bloch bands from higher excited state. Under this approach the second
%WS ladder states are very much unstable than the lowest one, therefore, we define
%the jump operators as $\hat C_{l}^b=\sqrt{\gamma_{l}^b} \hat b_l$. Within the nHA, 
%the norm of wavefunction is not preserved since ${\rm Tr}[\hat\rho_S]<1$. 
%The effective Hamiltonian will then preserve just total particle number (it decouples
%from Fock states with different total particle number) and not the energy. The reservoir 
%affects the system by inducing a dynamical dephasing via decay rates and thus driving 
%the system to an assymptotic state (steady state), our central interest in this work. 

Then our system Hamiltonian $\hat H$ for two-bands can be written as \cite{parraphd, Comp2015}
\begin{eqnarray}\label{eq:03}
\hat H &=& \sum_{\alpha={\{a,b\}}}\sum_l J_{\alpha}(\hat\alpha^{\dagger}_{l+1}\hat \alpha_l+{\rm h.c.})+
\frac{U_{\alpha}}{2} \hat \alpha^{\dagger 2}_l \hat{\alpha}_l^2+\varepsilon_l^{\alpha}\hat n^{\alpha}_l\notag\\
&+& \sum_{\mu = \{0,\pm 1\}}\sum_l c_{\mu} F (\hat b^{\dagger}_{l+\mu}\hat a_l+{\rm h.c.})\notag\\
&+&\sum_l \frac{U_x}{2} (\hat b^{\dagger 2}_l\hat a_l^2+{\rm h.c.})+2U_x \hat n^{a}_l\hat n^{b}_l \,,
\end{eqnarray}
with $\varepsilon_l^{\alpha}=d_LFl+\Delta_{\alpha}$. 
$\Delta_{\alpha}=(\delta_{\alpha,b}-\delta_{\alpha,a})\Delta/2$ is the on-site 
energy splitting between the single-site modes. To simplify the notation, we set $a\equiv\alpha_1$ and 
$b\equiv\alpha_2$. $\delta_{i,j}$ is the Kronecker delta.
The  system parameters are: the hopping matrix elements $J_{a,b}$, the atom-atom interaction constants 
$U_{x,a,b}$, and the dipole coupling matrix elements $c_{\mu=0,\pm1}$ between the bands induced by the force. 
The lattice constant $d_L$ and $\hbar$ are set to one. 
The control parameter is the Stark force $F(t)$ which will be explicitly time-dependent. 
This allows us to study the response of the system to any specific protocol for the driving. 
Here, we restrict to a periodic driving $F(t)=A \cos(\omega t)$.
     
%To solve the differential equation Eq.~\ref{eq:03} for $\hat\rho_S$ without the last r.h.s term 
%is equivalent to solve the Schr\"odinger equation $i\partial_t|\psi_t\rangle=\hat H_{\rm eff}|\psi_t\rangle$, due to the preservation of the total particle number. This is unexpensive from the 
%computational point of view, and the nHA evolving state is always pure. We can infer  dynamical
%properties from  the quantum spectrum of $\hat H_{\rm eff}$, a very powerful tool to predict and
%design protocols. 

We work within the reduced Fock basis $|\{n\}\rangle\equiv|n^a_1n^a_2\cdots n^a_L,n^b_1n^b_2\cdots n^b_L\rangle$, 
with dimension $\mathcal N_f=(N+2L-1)/N!(2L-1)$ given $N$ bosonic atoms in a system 
with $L$ lattice sites. In the rest of this work, we consider unitary filling with $N=L$. 
The time evolution is numerically performed using a vector-optimized fourth-order Runge-Kutta routine
presented and benchmarked in Ref.~\cite{Comp2015}.

The aim of this paper is to study the interplay between dissipation and the periodic 
driving. It will be shown that the combination of this two processes reveals surprising
dynamical features of our system. For simplicity, we set 
$\gamma_b\equiv\gamma_l^b$ and therefore the system-environment copuling 
term is: $-i\gamma_b \sum_l\hat{b}_l^\dagger\hat{b}_l=-i\gamma_b \sum_l\hat n^b_l$. 
This latter condition is not really relevant since our results do not strongly depend 
on how the single-site decay rates are distributed. Our figure of 
merit is the expected value of inter-WS ladder population inversion operator 
\begin{equation}\label{eq:04}
W(t) =  \langle \psi_t| \sum\nolimits_l(\hat n^b_l-\hat n^a_l)|\psi_t\rangle
/\langle\psi_t|\psi_t\rangle\,.
\end{equation} 
We restrict to initial states of the type 
$|\psi_0^1\rangle =\sum\nolimits_{\{n\}} D_{\{n\}}^0|n^a_1n^a_2\cdots,000\cdots\rangle$, for which
all particles are in the lowest WS ladder at $t=0$. 
An example of this kind of states is the Mott-insulator state
represented by $|\psi^1_0\rangle = |111\cdots,000\cdots\rangle$, which is routinely prepared
in many experiments as those Refs~\cite{greiner1,BDZ2008}. 
Other types of initial conditions can be prepared taking 
as a starting point the Mott state. Examples of such different states are so-called doublon states \cite{Innsbruck2013, kolovskypra}
$|\psi_0^{2,3}\rangle=\{|2020\cdots,000\cdots\rangle,|0202\cdots,000\cdots\rangle\}$, which can be used
to simulate quantum magnetism with ultracold atoms \cite{greiner2}.
As will be shown later in this work, the precise choice of the initial state is not crucial,
since the structure of the dynamically generated states, if they exist, will turn out to be 
robust with respect to imperfections in the initial conditions.

\subsection{Resonant Manifolds}
\label{sec:2-2}

Following Ref. \cite{kolovskypra}, we transform the Hamiltonian in Eq.~\eqref{eq:03} into 
the rotating frame with respect to the operator $\hat A=\sum_{\alpha,l}(\Delta_{\alpha}-\frac{1}{2}U_{\alpha}+lF)\hat n^{\alpha}_l$. 
The single-site annihilation (creation) operators transform as
$$\hat\alpha_l\rightarrow \hat \alpha_l\exp[-i(\Delta_{\alpha}-\frac{1}{2}U_{\alpha})t-il\,\theta(t)]\,,$$
with $\theta(t)=\int_0^tF(t')dt'$, then after plugging it into the Hamiltonian, we obtain
\begin{eqnarray}\label{eq:05}
\hat H' &=& \sum_{\alpha={\{a,b\}}}\sum_l J_{\alpha}\hat\alpha^{\dagger}_{l+1}\hat \alpha_le^{-i\theta(t)}\notag\\
&+& \sum_{\mu,l} \tilde{c_{\mu}} [e^{-i((\omega+\Delta) t -\mu\theta(t))}+e^{-i((\Delta-\omega) t -\mu\theta(t))}]\hat b^{\dagger}_{l+\mu}\hat a_l\notag\\
&+&\sum_l \frac{U_x}{2} \hat b^{\dagger 2}_l\hat a_l^2e^{-i 2\Delta t}+E_{\mu}^{\lambda}[\{n\}]+{\rm H.c.}
\end{eqnarray}
where we have defined $\tilde{c}_{\mu}\equiv c_{\mu}A/2$ and the unperturbed energies as
$E_{\nu}^{\lambda}[\{n\}] = U(\nu+\lambda)$, with $\nu=\frac{1}{2}\sum\nolimits_l (n^b_l-n^a_l)^2$, 
$\lambda=3\sum\nolimits_ln^a_ln^b_l$.

\begin{figure}[t]
\includegraphics[width=0.9\columnwidth]{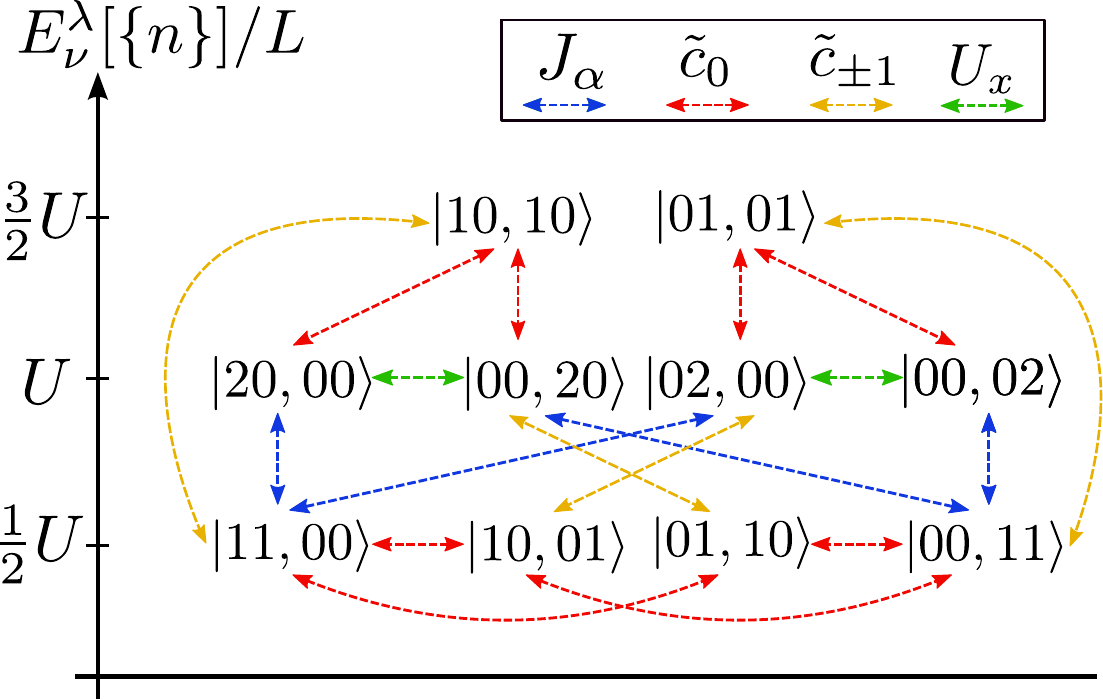}
\caption{\label{fig:reson} ({\em Color online}). Energy diagram and
allowed couplings in the many-body system for $N/L=2/2$, i.e., 
for two particles in two lattice sites. The color of the arrow defines the process, 
i.e. induced by hopping (blue), by the force (red and yellow), or by the interactions (green).}
\end{figure}

To simplify the discussion, we assume equal interaction strengths $U=U_a\approx U_{b}\approx U_{x}$. 
This approach is valid under the condition $J_{\alpha}\ll U_{a,b,x}$ as shown in \cite{parraphd}. 
The definition of $E_{\nu}^{\lambda}[\{n\}]$ permits us to split the Fock basis into sets of states with the same energy that 
can be exploited to enhance the inter-WS ladder coupling via resonant tunneling. The lowest-energy
manifold is that for $\nu=L/2$ and $\lambda=0$ corresponding to Fock states with
$n_l^a+n_l^b=1$. In Fig.~\ref{fig:reson}, we show the energy diagram for the case $N/L=2/2$ 
with the respective inter and intra-WS ladders coupling between the different states. 
Note that the pairs of states $\{|10,01\rangle,|01,10\rangle\}$,
$\{|20,00\rangle,|00,02\rangle\}$ and $\{|01,01\rangle,|10,10\rangle\}$ are not directly coupled
through the Hamiltonian terms. We will see that the principal effect of the combination 
of shaking and the dissipation is to 'freeze' the system into a certain linear 
combination of the first pair of states, which satisfies the condition 
$\sum_l (\hat n^b-\hat n^a_l)=0$. 

The transformed Hamiltonian $\hat H'$ has three different time scales related to every 
time-dependent term in Eq.~\eqref{eq:05}. These scales are defined by the periods $T_{\omega}=2\pi/\omega$, 
$T_{\Delta}=\pi/\Delta$, and $T^{\pm}_{\mu\neq 0}\approx 2A|\mu|/\omega(\Delta\pm \omega)$. 
For typical experimental parameters \cite{parraphd}, $\Delta$ is the largest energy scale. Therefore we consider 
frequencies with $\Delta> \omega$, and then $T_{\omega}> T_{\Delta}\gg T^{\pm}_{\mu\neq 0}$. 
This implies that the dynamical effects of the fast oscillating processes with frequencies 
$\omega_{\Delta}$, $\omega_{\mu\neq 0}^{\pm}$ happen within one period of the driving $F(t)$, 
and that they do not dominate the long-time evolution. 
%Yet, despite the steady state only appears for $t\ggg T_{\omega}$,  the Hamiltonian $\hat H'$
%is still time-dependent and periodic.

\section{Dissipation-induced asymptotic Dynamics and Entanglement generation}
\label{sec:3}

\begin{figure}[t]
\includegraphics[width=\columnwidth]{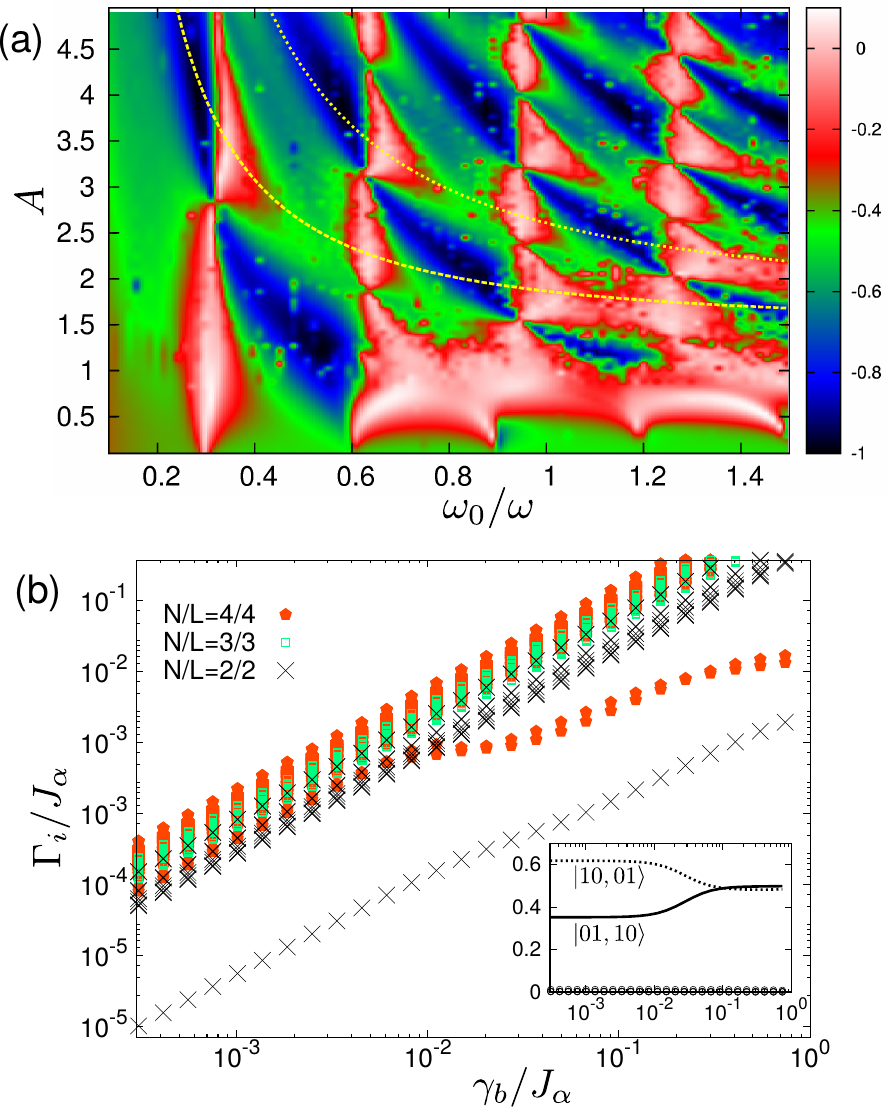}
\caption{\label{fig:1} ({\em Color online}). (a)
Long-time average of the population inversion $W(t)$ as a function of the 
amplitude $A$ and frequency $\omega$ of the drive $F(t)$. The 
initial condition is the Mott state $|11,00\rangle$. Here $\omega_0=\Delta-U$ 
and $\gamma_b=0.08 J_{\alpha}$, with $J_{\alpha}\equiv |J_{a,b}|$.
(b) Decay rates computed from the quasi-energyspectrum of $\hat H'$ for different
system sizes $N/L$ vs. the dissipation strengths (see legends). The inset shows the 
projections of the Fock states onto the Floquet eigenstates for $N/L=2/2$. 
The most stable state corresponds to a linear combination of only two Fock states 
belonging to the lowest resonant manifold.
The parameters  taken from \cite{parraphd} are: 
$\Delta= 0.1$, $J_a= -0.006$, $J_b = 0.006$, $U_a= U_b= U_x= 0.034$, 
$c_0 = -0.098$, and $c_{\pm}= 0.035$.}
\end{figure}

We compute the long-time average $\overline{W(t)}=\lim_{T\rightarrow \infty}T^{-1}\int_0^{T}W(t)dt$ 
in order to analyze the dependence of the inter-WS ladder dynamics on the driving and the
dissipation. The results are shown in the density plot in Fig.~\ref{fig:1}-(a)
for $\gamma_b/J_{\alpha}=0.08$ and $t\gg 2000 T_{\omega}$. This figure reveals the existence of 
resonance structures, i.e., finite-size areas in the parameter space at 
which the interband coupling saturates to $\overline{W(t)}\approx 0$, see the white areas in 
Fig.~\ref{fig:1}-(a). The regions are around discrete values for the frequency 
$\omega_n\propto \omega_0/n$, with $n\in \mathbb N$ with amplitude 
$A_n\propto 1/\sqrt{\omega_0/\omega}\propto \sqrt{n}$ in the strong driving regime, i.e. for
$\tilde{c}_{\mu}\gg J_{\alpha}$ (see yellow-dashed lines in Fig.~\ref{fig:1}-(a)). 
For weak driving, i.e., $\tilde{c}_{\mu}\sim J_{\alpha}$, the equidistance 
resonance structure no longer appears. Instead, only one area is found in the high frequency 
regime around $\omega_0/\omega \approx 1/3$. In the low frequency regime, there is a large region in the vicinity of $A=1$. 
%Despite the various resonance islands, 
%the results presented next are not sensible to the choice of island since for all of them 
%since we always have $\overline{W(t)}\approx 0$. The only chaning feature is the time scale 
%at which the steady state occurs. The fact that 
We learn from  Fig.~\ref{fig:1}-(a) that the probability of hitting a resonant structure is rather large. 
Moreover, the only difference between them is the time scale along which $W(t)\rightarrow 0$ for $t \gg T_{\omega}$.
$W(t)= 0$ means that the asymptotic state corresponds to an equal-population state, for
which the particle exchange between the two WS-ladders is frozen.

In order to understand the origin of the asymptotic state, we study the quasi-energy spectrum 
of $\hat H'$. We diagonalize the Floquet operator, i.e. the one-cycle evolution operator $\hat U(T_{\omega})$, and plot the
imaginary parts of the quasi-energies $\varepsilon_i = E_i-i\Gamma_i/2$. 
Fig.~\ref{fig:1}-(b) presents the decay rates $\Gamma_i$ as a function of 
the ratio $\gamma_b/J_{\alpha}$. Only if $N$ is even, highly stable states exist
with small decay rates $\Gamma_i$. For $N$ odd, we do not observe equilibration since no stable states are found. 
For $N/L=2/2$, the stable state is always found and can be expressed as a linear combination of the states 
$\{|10,01\rangle,|01,10\rangle\}$. Here, the individual weights become equal when 
increasing the dissipation strength. In the case $N/L=4/4$, the stable state
only appears when increasing the dissipation strength up to a critical value, 
which must be sufficiently strong to influence the long-time dynamics.
 
Given an initial state $|\psi_0^i\rangle$, the asymptotic dynamics is driven by the 
non-hermitian part of $\hat H_{\rm eff}$. 
To characterize the system's steady state, if it exists, we
write it in Fock basis representation as
\begin{eqnarray}\label{eq:06}
|\psi_{t\ggg T_{\omega}}\rangle=\sum_{\{n\}} D_{\{n\}}^t 
|n^a_1n^a_2\cdots n^a_L,n^b_1n^b_2\cdots n^b_L\rangle\,.
\end{eqnarray}
This permits us to study the asymptotic normalized probability distribution 
$p(\{|D_{\{n\}}^t|^2\})$ for different system sizes. 

Let us start with the minimal system $N/L=2/2$. The $|\psi_{t\gg T_{\omega}}\rangle$ 
results in the expected equally-weighted linear combination of the Fock states 
$\{|10,01\rangle,|01,10\rangle\}$ predicted in Fig.~\ref{fig:1}-(b) for
$\gamma_b/J_{\alpha}\approx 0.08$. Then we end up in the state
\begin{equation}\label{eq:07}
|\psi_{t\gg T_{\omega}}\rangle\approx \frac{1}{\sqrt{2}}(|01,10\rangle+e^{i\phi}|10,01\rangle)\,,
\end{equation}
which resembles a Bell state. The connection to the Bell basis is straightforward if we 
use instead the single-site population inversion number $w_l\equiv n^b_l- n^a_l$ 
such that we map the configuration $|n^a_1n^a_2\cdots n^a_L,n^b_1n^b_2\cdots n^b_L\rangle$ onto 
$|w_1w_2\cdots w_L\rangle$. Note that, for $|\psi_{t\gg T_{\omega}}\rangle$ in Eq.~\eqref{eq:07}, 
$w_l$ can only take the values $\pm 1$ and satisfies
$\langle \psi_{t\ggg T_{\omega}}|\sum\nolimits_l\hat w_l|\psi_{t\ggg T_{\omega}}\rangle \rightarrow 0$.
We introduce the pseudo-spin representation $w_l=-1_l \equiv \downarrow_l$ and $w_l=1_l \equiv \uparrow_l$, 
with which the state renders the Bell state 
$\frac{1}{\sqrt{2}}(\ket{\uparrow\downarrow} - \ket{\downarrow\uparrow}) \equiv |\Psi^-\rangle$. The phase factor 
$\phi \approx \pi$  is extracted from the numerics.

\begin{figure}[t]
\includegraphics[width=0.83\columnwidth]{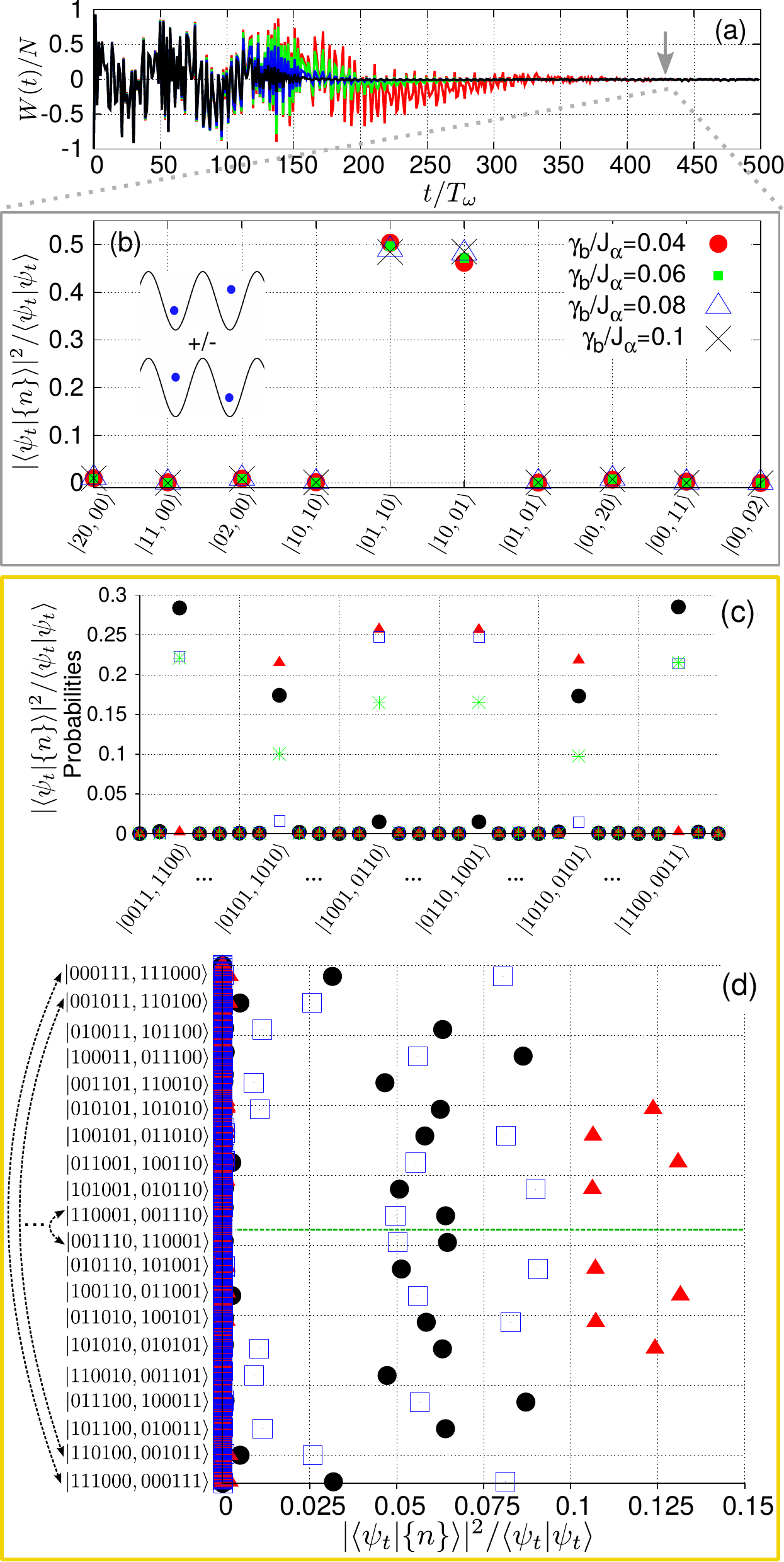}
\caption{\label{fig:2} ({\em Color online}) (a) and (b) show
$W(t)$ for $N/L=2/2$ and the final-time normalized distribution $p(\{|D^t_{\{n\}}|^2\})$  
for the asymptotic state vs. the dissipation strength. 
The initial condition is the Mott-insulator state defined in main text. Panels (c) and (d)
show $p(\{|D^t_{\{n\}}|^2\})$ for $N/L=4/4$ and $N/L=6/6$.
Initial conditions: (black-filled circles) $\ket{\psi_0^1}$, (blue-squares) $\frac{1}{\sqrt{2}}(\ket{\psi_0^2}+\ket{\psi_0^3})$
and (red-filled triangles) $\frac{1}{\sqrt{2}}\ket{\psi_0^1}+\frac{1}{2}(\ket{\psi_0^2}+\ket{\psi_0^3})$ defined in the main text.
The green asterisks in the panel (c) correspond to an average
over 100 initial random states of the type $\sum D_{\{n\}}|n^a_1n^a_2\cdots,00\cdots\rangle$.
Note that the states of interest satisfy $\sum\nolimits_l(\hat n^b_l-\hat n^a_l)=0$
and $n^a_l+n^b_l=1$. These conditions imply a band-exchange symmetry of the populations discussed
at the end of section \ref{sec:3}. The parameters are the same of Fig.~\ref{fig:1} with $|c_{\mu}A/J_a|\simeq 10$ 
and $\omega=\omega_0$.}
\end{figure}

Remarkably, the results for $N/L=2/2$ can be extended to larger systems. 
The probability distributions $p(\{|D_{\{n\}}^t|^2\})$ for  $N/L=\{4/4; 6/6\}$
are shown  in Fig.~\ref{fig:2}-(c,d). Again, pairs of equally-weighted
states have the largest contributions. For $N/L=4/4$ the pseudo-spin representation 
for $|\psi_{t\gg T_{\omega}}\rangle$ is given by
\begin{eqnarray}\label{eq:08}
|\psi_{t\gg T_{\omega}}\rangle &\approx & |D_1^t| (|\uparrow\uparrow\downarrow\downarrow\rangle
+e^{i\phi_1}|\downarrow\downarrow\uparrow\uparrow\rangle)\notag\\
&+& |D_2^t| e^{i\theta_1^2}(|\uparrow\downarrow\uparrow\downarrow\rangle+e^{i\phi_2}|\downarrow\uparrow\downarrow\uparrow\rangle)\notag\\
&+& |D_3^t| e^{i\theta_1^3}(|\downarrow\uparrow\uparrow\downarrow\rangle+e^{i\phi_3}|\uparrow\downarrow\downarrow\uparrow\rangle)\,.
\end{eqnarray}
It satisfies $w_l \approx \pm 1$ and $W(t\gg T_{\omega})\approx 0$. From the numerics, 
the phase differences are found this time to be $\phi_i\approx 0$. A Bell-like state structure is recovered after rewriting
$|\psi_{t\gg T_{\omega}}\rangle$ using a set of new labels based on the spatial lattice bipartition with 
respect to the center of the lattice
\begin{equation}\label{eq:09}
|\{w\}\rangle = |\underbrace{w_1w_2\cdots w_{L/2}}_\text{Left}\rangle\otimes|\underbrace{w_{L/2+1}\cdots w_{L}}_\text{Right}\rangle\,.
\end{equation}

A careful analysis of the structure of the asympotic state in the pseudo-spin representation 
suggests the following definitions  
$$d\equiv \downarrow_1\downarrow_2\,,\,\,\,\bar{d}\equiv \uparrow_3\uparrow_4\,,\,\,\,o\equiv \downarrow_1\uparrow_2\,,\,\text{and}\,\,\,\bar{o}\equiv \uparrow_3\downarrow_4\,,$$ 
with which the state gets the form
\begin{eqnarray}\label{eq:10}
|\psi_{t\ggg T_{\omega}}\rangle &\approx & |D_1^t| (|d\bar{d}\rangle+|\bar{d}d\rangle)\notag\\
                   &+& |D_2^t| e^{i\theta_1^2}(|o\bar{o}\rangle+|\bar{o}o\rangle)\notag\\
                   &+& |D_3^t| e^{i\theta_1^3}(|oo\rangle+|\bar{o}\bar{o}\rangle)\notag\\
             &\approx & |D_1^t| |\Psi^+_d\rangle+ |D_2^t| e^{i\theta_1^2}|\Psi^+_o\rangle
         + |D_3^t| e^{i\theta_1^3}|\Phi^+_o\rangle\,.\notag\\
\end{eqnarray}
Hence, $|\psi_{t\gg T_{\omega}}\rangle$ can be mapped onto a linear 
combination of the Bell states $|\Psi^+_x\rangle\equiv \frac{1}{\sqrt{2}}(\ket{x \bar{x}} + \ket{\bar{x}x\rangle}$
and  $|\Phi^+_x\rangle\equiv \frac{1}{\sqrt{2}}(\ket{xx} + \ket{\bar{x}\bar{x}\rangle}$,
taking into account the new set of labels $x\in\{d,\bar{d},o,\bar{o}\}$. Therefore, spatial entanglement between the two
lattice subsystems persists when increasing the system size.

As anticipated, the structure of the asymptotic state does not depend much
on the initial state, see Fig.~\ref{fig:2}-(c). We observe a dependence of 
the weights of the relevant states on the initial state choice, though the 
symmetry is preserved. This is verified after averaging over randomized linear 
combination of lower band Fock states: the asymptotic
distribution $p(\{|D_{\{n\}}^t|^2\})$ preserves the above Bell-state like 
structure. The time scale for the formation of the asymptotic state depends on 
the system-environment coupling strength and the system size \footnote{For the
systems $N/L=4/4$ and $N/L=6/6$ the asymtotic states occurs for $t\sim 2000 T_{\omega}$ 
and $t\sim 8000 T_{\omega}$, respectively given $\gamma_b=(0.08-0.1) J_{\alpha}$.}, see discussion at 
the beginning of this section. This time scale depends on the norm weight of the 
basin of attraction of the asymptotic states of interest here.
Therefore, the asymptotic state structure is found to be very robust with 
respect to imperfections in the initial state preparation. 

In Fig.~\ref{fig:2}-(d) we show the probability distribution for $N/L=6/6$ given
the initial states $|\psi_0^i\rangle$. Despite the large dimension of the 
associated Fock space, we immediately notice that the relevant contributions 
in the asymptotic state come, again, by pairs, as in the previous cases. 
Following transformations as described above, we have
\begin{eqnarray}\label{eq:11}
|\psi_{t\gg T_{\omega}}\rangle &=& |D_1^t| (|d\bar{d}\rangle+|\bar{d}d\rangle)\notag\\
&+&|D_2^t|e^{i\theta_1^2}(|sp\rangle+|ps\rangle)\notag\\
&+&|D_3^t|e^{i\theta_1^3}(|\bar{o}p\rangle+|p\bar{o}\rangle)\notag\\
&+&|D_4^t|e^{i\theta_1^4}(|\bar{p}p\rangle+|p\bar{p}\rangle)\notag\\
&\vdots& \notag\\
&+&|D_{10}^t|e^{i\theta_1^{10}}(|s\bar{s}\rangle+|\bar{s}s\rangle)\,.
\end{eqnarray}
with 
\begin{eqnarray}\label{eq:12}
d\equiv\downarrow_1\downarrow_2\downarrow_3,\,\,\bar{d}\equiv\uparrow_4\uparrow_5\uparrow_6 
&& o\equiv\downarrow_1\uparrow_2\downarrow_3,\,\,\bar{o}\equiv\uparrow_4\downarrow_5\uparrow_6\notag\\
s\equiv\downarrow_1\downarrow_2\uparrow_3,\,\,\bar{s}\equiv\uparrow_4\uparrow_5\downarrow_6 
&& p\equiv\uparrow_1\downarrow_2\downarrow_3,\,\,\bar{p}\equiv\downarrow_4\uparrow_5\uparrow_6\notag \,.
\end{eqnarray}
We see that again a sort of superposition of Bell states is recovered. 

In order to confirm that the states for larger systems are indeed spatially entangled  
we compute the entanglement negativity $E_{\mathcal N}(\hat\rho_t)=(\|\rho^{\Gamma_{\mathcal R}}\|_1-1)/2$ 
\cite{Luba2011,negativity}. It is
defined via the partial transpose of the density operator $\hat\rho^{\Gamma_{\mathcal R}}$ taken over
right-side subsystem partition $\mathcal R$, whose matrix elements in the $w_l$ 
representation are
\begin{eqnarray}\label{eq:13}
\langle\{s\}|\hat\rho_t^{\Gamma_{\mathcal R}}|\{v\}\rangle = \sum_{\{w\},\{w'\}} D_{\{w\}}^tD_{\{w'\}}^{t\,*}
\prod_{l=1}^{L/2}\delta_{s_l,w_l}\delta_{w_l',v_l}\notag\\
\times\prod_{l=L/2+1}^{L}\delta_{s_l,w_l'}\delta_{w_l,v_l}\,.
\end{eqnarray}

\begin{figure}[t]
\includegraphics[width=\columnwidth]{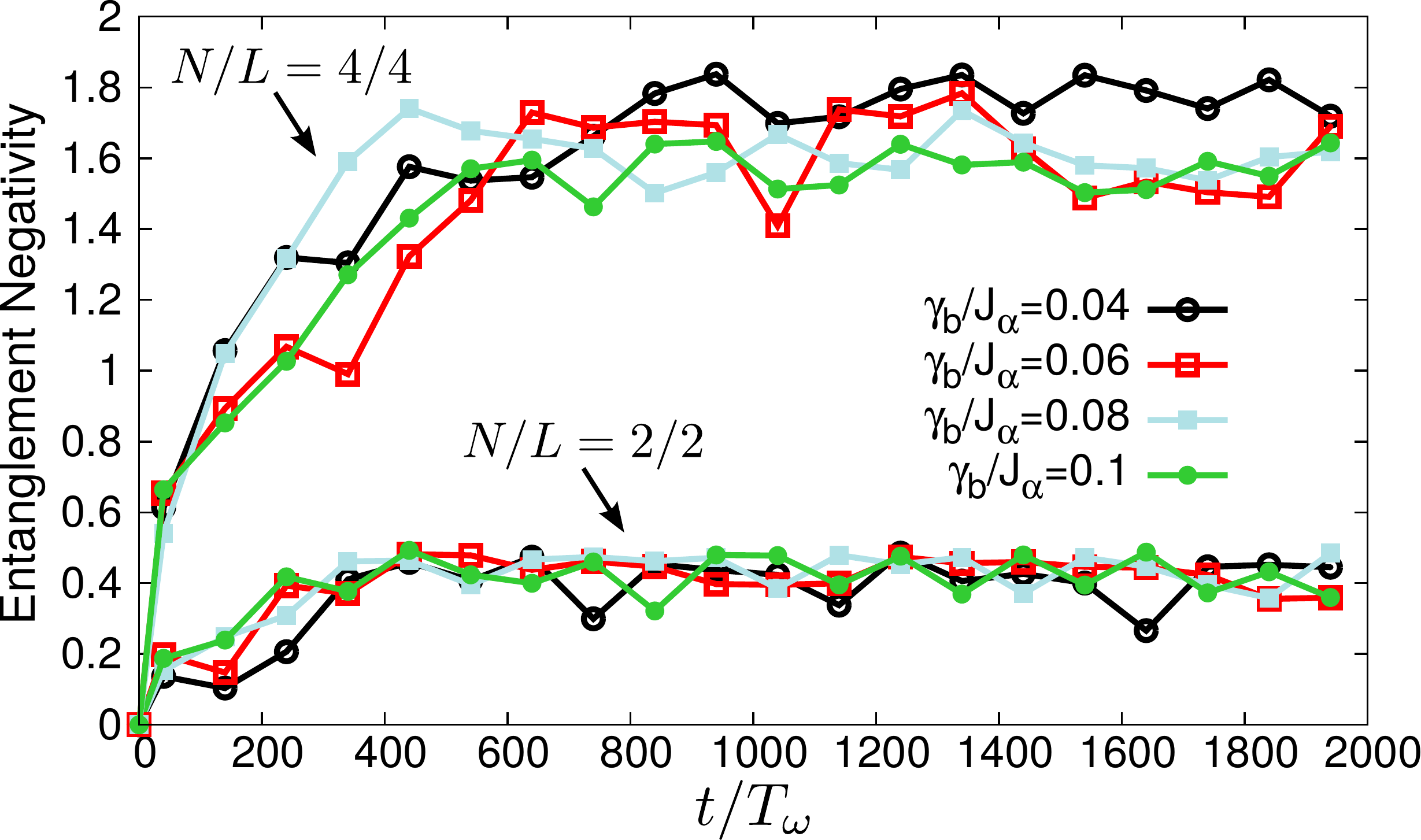}
\caption{\label{fig:3} ({\em Color online}) Entanglement negativity for $N/L=2/2$ and $N/L=4/4$
for increasing dissipation $\gamma_b/J_{\alpha}$. The initial state is a non-entangled Mott-insulator state. 
The parameters are the same as in Figs.~\ref{fig:1} and \ref{fig:2}.}
\end{figure}

It is known that $E_{\mathcal N}(\hat\rho_t)$ is an entanglement witness, i.e. it only detects entanglement 
rather than quantifying it. The matrix is numerically diagonalized using LAPACK subroutines 
to obtain the eigenvalues. We evaluate the trace norm as
$\|\hat\rho^{\Gamma_{\rm R}}\|_1=\sum_i|\lambda_i|$. In Fig.~\ref{fig:3}, we show the 
onset of entanglement between the subsystem partitions ${\mathcal L}$eft and ${\mathcal R}$ight 
of the lattice, see also \cite{Luba2011} where a similar partition was chosen. 
The figure confirms that, for larger systems, the evolving state gets entangled in time.
Moreover, it also shows a saturation exactly at the interband equilibrium, i.e., when the system steady state regime is reached.
For the case $N/L=2/2$, this is seen in comparison with Fig.~\ref{fig:2}(a).

The previous analysis of the asymptotic dynamics reveals that the system is driven into a
quasi loss-free subspace set by the following conditions: 
$(i)$ $n_l^{a}+n_l^{b}=1$ (or equivalently $w_l=\{-1,1\}$), and $(ii)$ $\sum\nolimits_l w_l = 0$.
These conditions are satisfied only for a small subset of states from the 
low-energy manifold $E^{L/2}_0$  which are not directly coupled via the Hamiltonian 
terms in Eq.~(\ref{eq:05}) (see Fig.~\ref{fig:reson}). The structure of the asymptotic state implies the existence of 
a subspace of states $\mathcal S$ formed by pairs of Fock states, the linear combination of which is invariant 
under the exchange of the on-site populations between the two bands, for example,
$\sim|\uparrow\downarrow\uparrow\downarrow\cdots\rangle \pm |\downarrow\uparrow\downarrow\uparrow\cdots\rangle$. 
Whether the state is symmetric or anti-symmetric the condition $(ii)$ is always satisfied (see inset in Fig.~\ref{fig:2}-(b)).

In addition, the asymptotic state never has more than one atom per lattice 
site. Therefore, it is "free" of interaction effects, and we might operate on it via the dipole $c_{0}$ term without destroying 
its symmetry $(i,ii)$, with an appropriate choice of the drive $F(t)$.

The symmetry above exposed is broken if $N$ is an odd number since the internal subspace $\mathcal S$
can be split into two subspaces with $\sum w_l=\pm 1$. Yet if the evolving state can be written as a linear 
combination of pairs of states with the mirror symmetry we have
$\langle \psi_{t\ggg T_{\omega}}|\sum\nolimits_l\hat w_l|\psi_{t\ggg T_{\omega}}\rangle \approx 0,\pm 2$, i.e., 
there is no privileged value for the inter-WS ladder exchange and the system 
will not equilibrate. This is the reason why we observe no stable state in the quasi-spectrum analysis in Fig.~\ref{fig:1}-(b).
$N$ odd also affects the definition of the spatial bipartition that generates entanglement and the Bell-like
structures since there is no geometrical center which allows for the introduction of the labels $x$. 
Therefore the spatial entanglement, if possible, is far more complicated to study. For instance, if we
set periodic boundary conditions there is no geometrical center and we cannot straightforwardly define spatial bipartitions 
as in Eq.~(\ref{eq:09}).

\section{Conclusions}
\label{sec:concl}

We have investigated a two-band many-body Wannier-Stark model with an interplay 
between periodic driving and an effective dissipative coupling to higher bands. This is a first study combining
previous work on closed two-band Bose-Hubbard models  \cite{parraphd, Comp2015} with open single-band many-body
problems \cite{EPJ2015}. 

The driven many-body system shows remarkable features such as the dynamical generation of entanglement and the evolution 
into dynamically decoupled asymptotically stable states. These states obey a band-population exchange symmetry that is responsible for the creation of maximally entangled Bell-like states that appear after a convenient reorganization of their Fock representation. 
The spatial 'left-right' entanglement is detected by means of the negativity as an entanglement witness. Our findings turn out to be sensitive only with respect to the precise geometry of the system, i.e., they occur for a lattice with an even number of sites and hard-wall boundary conditions.

Our results are, as shown, robust to imperfections in the preparation of the initial state, which gives good 
perspectives for future applications in experiments with ultracold atoms \cite{Entangle,Greiner2015,Bloch2015} and particularly in 
quantum information processing \cite{lidar}.

\section{Acknowledgments}
The authors acknowledge financial support of the University del Valle (project CI 7996).
C. A. Parra-Murillo gratefully acknowledges financial support of COLCIENCIAS (grant 656) and
SW support by the FIL2014 program of Parma University. We warmly thank J. H. Reina and C. Susa for lively discussion. 
Moreover, we are very grateful to Elmar Bittner for computational and logistic support at the ITP in Heidelberg.

\end{document}